\documentclass[prd,nofootinbib,floats,aps,superscriptaddress,twocolumn]{revtex4}

\usepackage{hyperref}
\usepackage{graphicx}
\usepackage{slashed}
\usepackage{amsmath,amssymb}
\usepackage{hhline}

\newcommand{\beq}{\begin{equation}}
\newcommand{\eeq}{\end{equation}}
\newcommand{\beqa}{\begin{eqnarray}}
\newcommand{\eeqa}{\end{eqnarray}}

\newcommand{\Bbar}{\,\overline{\!B}{}}
\newcommand{\Dbar}{\,\overline{\!D}{}}
\newcommand{\Kbar}{\,\overline{\!K}{}}
\def\B0bar{\Bbar{}^0}
\def\D0bar{\Dbar{}^0}
\def\K0bar{\Kbar{}^0}

\begin{document}

\title{\boldmath Lepton flavor-violating transitions in effective field theory and 
gluonic operators}

\def\wsu{Department of Physics and Astronomy, 
Wayne State University, Detroit, MI 48201, USA}
\def\mctp{Michigan Center for Theoretical Physics, 
University of Michigan, Ann Arbor, MI 48109, USA}

\author{Alexey A.\ Petrov}
\affiliation{\wsu}\affiliation{\mctp}
\author{Dmitry V.\ Zhuridov}
\affiliation{\wsu}

\date{\today}


\begin{abstract}
Lepton flavor-violating processes offer interesting possibilities to probe new physics at multi-TeV scale. 
We discuss those in the framework of effective field theory, emphasizing the role of gluonic operators.
Those operators are obtained by integrating out heavy quarks that are kinematically inaccessible 
at the scale where low-energy experiments take place and make those experiments sensitive to 
the couplings of lepton flavor changing neutral currents to heavy quarks. 
We discuss constraints on the Wilson coefficients of those operators from 
the muon conversion  $\mu^- + (A,Z) \to e^- + (A,Z)$ and from lepton flavor-violating tau 
decays with one or two hadrons in the final state, e.g. $\tau \to \ell \ \eta^{(\prime)}$ and 
$\tau \to \ell \ \pi^+\pi^-$ with $\ell = \mu, e$. To illustrate the results we discuss explicit examples 
of constraining parameters of leptoquark models.
\end{abstract}

\maketitle

\section{Introduction}

As follows from observations of neutrino oscillations, there is a good evidence that the individual lepton 
flavor is not conserved. Explicit calculations of the standard model (SM) rates for the charged lepton-flavor 
violating (LFV) transitions indicate that those are tiny~\cite{Marciano:1977wx,Vergados:1985pq}, well 
beyond the capabilities of current and currently planned experiments. Yet, many models of beyond the 
standard model (BSM) physics do not exclude relatively large rates for such transitions, so experimental 
and theoretical studies of LFV processes like $\mu \to e \gamma$, $\tau \to \eta^{(\prime)}\mu$ or
 $\mu^- + (A,Z) \to e^- + (A,Z)$ could provide a sensitive test of those BSM schemes.

The language of effective field theory (EFT) is very useful in the studies of flavor violating 
processes for several reasons. First, it allows to probe the new physics (NP) scale generically, 
without specifying a particular model of NP. Studies of specific models in this framework 
are equivalent to specifying Wilson coefficients of effective operators. Second, EFT allows for studies of 
relative contributions of various operators and may even provide clues as to what experiments 
need to be done to discriminate among different possible models of new physics~\cite{Cirigliano:2009bz}.  

Interactions of flavor-changing neutral currents (FCNC) of leptons with hadrons, either in muon conversion  
or in tau or meson decays can be described in terms of effective operators of increasing dimension~\cite{Cirigliano:2009bz}.
In order to set up an EFT calculation, however, one must first discuss the multitude of scales present in lepton 
FCNC transitions. The highest scale, which we denote as $\Lambda$, is the scale associated with new physics 
that generates the FCNC interaction. There could be many ways to generate the flavor-changing neutral 
current of leptons, yet, below the scale $\Lambda$ any heavy new physics particles are integrated out resulting only 
in a few effective operators~\cite{Golowich:2007ka}. We shall keep track of the leading contribution due to NP which, 
below the scale $\Lambda$, is proportional to $1/\Lambda^2$. The second highest scale is the one associated with 
electroweak symmetry breaking, $v$. The most important scales for this study are the scales associated 
with heavy quark masses, $m_t$, $m_b$, and $m_c$. In the framework of EFT one must integrate out heavy 
quarks that are not kinematically accessible at the scale where the experiment takes place, resulting in changes of 
Wilson coefficients of quark and gluon operators. 

The relation between all those scales can be done with the help of renormalization group, keeping track of which
degrees of freedom are kept and which are integrated out. We shall list the most important operators for our 
analysis below.

\subsection{Quark operators}

The lowest dimensional local operators that contribute to lepton-flavor violating transitions without 
photons in the final state~\cite{MuEGamma} have operator dimension 6. There are, in general, twelve types of operators 
that can be constructed,
\begin{eqnarray}\label{eq:L_eff_6}
 \mathcal{L}_{\rm \ell_1\ell_2}^{(6)} =  \frac{1}{\Lambda^2} \sum_{i=1}^{12} \sum_q C_i^{q\ell_1\ell_2} Q_i^{q\ell_1\ell_2}  + \text{H.c.},
\end{eqnarray}
where $\Lambda$ is a high scale of new physics, and $C_i^{q\ell_1\ell_2}$ are dimensionless 
Wilson coefficients. 
The four fermion operators can be split into three classes which we define according to 
their Dirac structure,

\noindent
(i) scalar operators, 
%
\begin{eqnarray}\label{Eq:ScalOps}
& & Q_1^{q\ell_1\ell_2} =  (\bar \ell_{1R} \ell_{2L}) \ (\bar q_R q_L), 
\nonumber \\
& &  Q_2^{q\ell_1\ell_2} =  (\bar \ell_{1R}\ell_{2L}) \ (\bar q_L q_R), 
\nonumber \\
& & Q_3^{q\ell_1\ell_2} =  (\bar \ell_{1L} \ell_{2R}) \ (\bar q_R q_L), 
\\
& & Q_4^{q\ell_1\ell_2} =  (\bar \ell_{1L} \ell_{2R}) \ (\bar q_L q_R), 
\nonumber
\end{eqnarray}
where $\ell$ ($q$) is the SM charged lepton (quark).

The scalar operators above are defined below the scale of electroweak symmetry breaking (EWSB) in the standard model
as they are not invariant under electroweak $SU(2)_L$-symmetry. The proper definition of those operators 
above EWSB scale should include Higgs doublet fields $H$. The operators of Eq.~(\ref{Eq:ScalOps}) result from 
the substitution $H \to v$ and redefinition of proper Wilson coefficients~\cite{ScalarOp} to scale out quark or lepton Yukawa coupling, 
which would result in (dimensionless) factor of $G_F m_\ell m_q$ in front of the scalar operators.

These mass factors properly suppress flavor-violating transitions of the first generation of quarks 
and leptons that are well constrained experimentally. Notice, however, that they are not model-universal. 
For example, models with FCNC Higgs boson interactions often employ factors of 
$\sqrt{m_{\ell_1} m_{\ell_2}}/v$ (so called Cheng-Sher ansatz~\cite{Cheng:1987rs}) 
to suppress flavor-changing lepton currents, while generic leptoquark or 
R-parity violative supersymmetric models do not have any factors of mass, relying on the smallness of coupling constants to 
suppress those effects \cite{Li:2005rr}. In the following we shall absorb all mass factors into the 
definition of Wilson coefficients $C_i^{q\ell_1\ell_2}$.

\noindent
(ii) vector operators,
\begin{eqnarray}\label{Eq:VecOps}
& & Q_5^{q\ell_1\ell_2} = (\bar \ell_{1L} \gamma^\mu \ell_{2L}) (\bar q_L \gamma_\mu q_L), 
\nonumber \\
& & Q_6^{q\ell_1\ell_2} = (\bar \ell_{1L} \gamma^\mu \ell_{2L}) (\bar q_R \gamma_\mu q_R), 
\nonumber \\
& & Q_7^{q\ell_1\ell_2} = (\bar \ell_{1R} \gamma^\mu  \ell_{2R}) (\bar q_L \gamma_\mu q_L), 
\\
& & Q_8^{q\ell_1\ell_2} = (\bar \ell_{1R} \gamma^\mu  \ell_{2R}) (\bar q_R \gamma_\mu q_R), 
\nonumber 
\end{eqnarray}
and (iii) tensor operators,
\begin{eqnarray}\label{TenOps}
& & Q_9^{q\ell_1\ell_2} = (\bar \ell_{1R} \sigma^{\mu\nu} \ell_{2L}) (\bar q_R \sigma_{\mu\nu} q_L),  
\nonumber \\
& & Q_{10}^{q\ell_1\ell_2} = (\bar \ell_{1R} \sigma^{\mu\nu} \ell_{2L}) (\bar q_L \sigma_{\mu\nu} q_R), 
\nonumber \\
& & Q_{11}^{q\ell_1\ell_2} = (\bar \ell_{1L}L \sigma^{\mu\nu} \ell_{2R}) (\bar q_R \sigma_{\mu\nu} q_L), 
\\
& & Q_{12}^{q\ell_1\ell_2} = (\bar \ell_{1L} \sigma^{\mu\nu} \ell_{2R}) (\bar q_L \sigma_{\mu\nu} q_R),
\nonumber
\end{eqnarray}
All quark flavors need to be considered, but the operator basis needed to describe a particular
experiment could include a smaller number of operators. 

\subsection{Gluonic operators}

The low energy experiments such as muon conversion $\mu+N \to e+N^\prime$ or tau decay
 $\tau \to \eta^{(\prime)}\mu$ have a naturally defined scale of the order of the
 mass of heavier lepton. In order to write appropriate set of effective operators at that scale one must
 integrate out quarks with masses above that scale~\cite{Potter:2012yv}. 

The flavor changing Lagrangian for the effective vertices with $\ell_1$, $\ell_2$ and two gluon 
external legs at the energies lower than heavy quarks masses can be 
written as
\begin{eqnarray}\label{eq:L_eff}
 \mathcal{L}_{\rm \ell_1\ell_2}^{(7)} =  \frac{1}{\Lambda^2} \sum_{i=1}^4 c_i^{\ell_1\ell_2} O_i^{\ell_1\ell_2}  + \text{H.c.},
\end{eqnarray}
where $c_i$ are the Wilson coefficients, and $O_i$ are the effective operators of dimension 7:
\begin{eqnarray}
& & O_1^{\ell_1\ell_2} = \bar \ell_{1R} \ell_{2L} \  \frac{\beta_L}{4\alpha_s} G_{\mu\nu}^a G^{a\mu\nu}, 
\nonumber \\ 
& & O_2^{\ell_1\ell_2} = \bar \ell_{1R} \ell_{2L} \  \frac{\beta_L}{4\alpha_s} G_{\mu\nu}^a \widetilde G^{a\mu\nu},
\nonumber \\ 
& & O_3^{\ell_1\ell_2}  = \bar \ell_{1L} \ell_{2R} \  \frac{\beta_L}{4\alpha_s} G_{\mu\nu}^a G^{a\mu\nu},  
\label{eq:Oprs}
\\ 
& & O_4^{\ell_1\ell_2} = \bar \ell_{1L} \ell_{2R} \  \frac{\beta_L}{4\alpha_s} G_{\mu\nu}^a \widetilde G^{a\mu\nu},  
\nonumber 
\end{eqnarray}
where $a=1,\dots,8$ is gluon color index, $\beta_L = -b\alpha_s^2/(2\pi)$ is the one-loop beta function of three flavor 
QCD with $b=11-2n_L/3$  ($n_L=3$ is the number of light quarks) and $\alpha_s=g_s^2/(4\pi)$;
\begin{eqnarray}
 G_{\mu\nu}^a = \partial_\mu A_\nu^a -\partial_\nu A_\mu^a + g_sf^{abc}A_{\mu}^bA_{\nu}^c
\end{eqnarray}
is the gluon strength tensor, and
\begin{eqnarray}
 \widetilde G_{\mu\nu}^a = \frac{\text{1}}{2} \varepsilon_{\mu\nu\alpha\beta}G^{a\alpha\beta}
\end{eqnarray}
is  the dual one. 
In Eq.~\eqref{eq:L_eff} we do not include the operators with dimension higher than 7. It can be easily seen 
that there are no other possibilities besides the four operators in Eqs.~\eqref{eq:Oprs}.

\begin{figure}
  \centering
   \includegraphics[width=.23\textwidth]{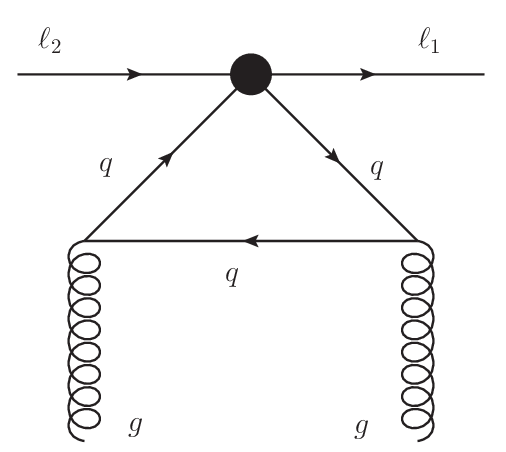}
   \includegraphics[width=.23\textwidth]{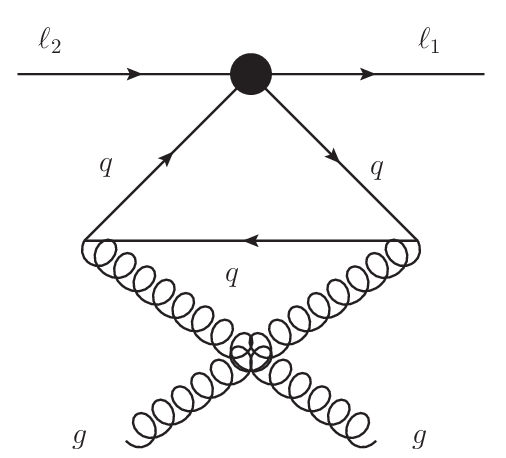}
   \caption{Feynman graphs for the calculation of matching coefficients of gluonic operators. The large dots indicate the effective vertices described by Eq.~\eqref{eq:L_eff_6}.}\label{Fig:1}
\end{figure}
By calculating the loop diagrams in Fig.~\ref{Fig:1}, using the standard methods~\cite{Peskin:1995ev}, the coefficients $c_i ^{\ell_1\ell_2} $ can be 
expressed through $C_i^{q\ell_1\ell_2}$ in Eq.~\eqref{eq:L_eff_6} as
\begin{eqnarray}
 	c_1^{\ell_1\ell_2}	&=&	-\frac{2}{9}	\sum_{q=c,b,t}	\frac{I_1(m_q)}{m_q}		(C_1^{q\ell_1\ell_2} + C_2^{q\ell_1\ell_2}),  \label{eq:c1}	\\
	c_2^{\ell_1\ell_2}	&=&	 \frac{2\text{i}}{9}	\sum_{q=c,b,t}	\frac{I_2(m_q)}{m_q}		(C_1^{q\ell_1\ell_2} - C_2^{q\ell_1\ell_2}),	\\
	c_3^{\ell_1\ell_2}	&=&	-\frac{2}{9}	\sum_{q=c,b,t}	\frac{I_1(m_q)}{m_q}		(C_3^{q\ell_1\ell_2} + C_4^{q\ell_1\ell_2}),  \label{eq:c3}	\\
	c_4^{\ell_1\ell_2}	&=&	 \frac{2\text{i}}{9}	\sum_{q=c,b,t}	\frac{I_2(m_q)}{m_q}		(C_3^{q\ell_1\ell_2} - C_4^{q\ell_1\ell_2}),  \label{eq:c4}
\end{eqnarray}
where the coefficients (see Ref.~\cite{Rizzo:1979mf} for $I_1$ and Appendix \ref{Ap:Integrals}) 
at the leading order are
\begin{eqnarray}\label{eq:I1result}
	I_1 =   \frac{1}{3}, \quad I_2 =   \frac{1}{2}
\end{eqnarray}
Note, as previously discussed, that while the Wilson coefficients in Eq.~(\ref{eq:c1})-\eqref{eq:c4} explicitly contain 
factors of $1/m_q$, in many models 
the coefficients $C_i^{q\ell_1\ell_2}$ contain factors of $m_q$, which we absorbed as part of their definition. 
Also, we do not explicitly write out contributions to Wilson coefficients due to possible colored heavy states that 
are not SM quarks; those contributions would result in additive modifications of 
Eqs.~(\ref{eq:c1})-\eqref{eq:c4}. Also, in this paper, we ignored running of $c_i$ in between 
different scales.

Integrating out heavy particles could also result in higher dimensional gluonic operators, as would happen for 
vector-like dimension-6 operators. For instance, a set of operators of dimension 8 can be written as
\begin{eqnarray}\label{eq:Vec_dim8}
O_1^{(8)} &=&  \epsilon^{\mu\nu\alpha\beta} 
(\bar \ell_{1L} \partial_\mu \gamma_\nu \ell_{2L})  \frac{\beta_L}{4\alpha_s} G^a_{\alpha\rho} G^{a\rho}_\beta \,
\nonumber \\
O_2^{(8)} &=&  \epsilon^{\mu\nu\alpha\beta} 
(\bar \ell_{1L} \partial_\mu \gamma^\rho \ell_{2L})  \frac{\beta_L}{4\alpha_s} G^a_{\rho\nu} G^a_{\alpha\beta} \,
\\
O_3^{(8)} &=& \qquad
~(\bar \ell_{1L} \partial_\mu \gamma^\rho \ell_{2L})  \frac{\beta_L}{4\alpha_s} G_{\rho\alpha} \widetilde G^{a\mu\alpha} \,
\nonumber
\end{eqnarray}
Another three operators, $O_4^{(8)}-O_6^{(8)}$ could be obtained by substituting left-handed
lepton fields with the right-handed ones. Here we shall concentrate on the operators of dimension 7, leaving 
analysis of higher-dimensional operator for future work.

This paper is organized as follows. In section~\ref{Ch:Conversion} we reexamine constraints on the Wilson coefficients
of operators $O_1^{e\mu}$ and $O_3^{e\mu}$ from $\mu$-$e$ conversion data. We consider constraints on Wilson coefficients of 
operators $O_1^{\ell\tau}-O_4^{\ell\tau}$ from tau decays in section~\ref{Ch:TauDecays}. As an example, in 
section~\ref{Ch:Leptoquarks} we consider how our constraints translate into constraints on couplings of LFV lepton currents 
with heavy quarks in leptoquark models. We conclude in section~\ref{Ch:Conclusions} .

\section{Constraints from $\mu$-$e$ conversion}\label{Ch:Conversion}

Muon conversion on a nucleus~\cite{Marciano:1977cj,Shanker:1979ap,Vergados:1985pq,Bernabeu:1993ta,Kosmas:1993ch,Barbieri:1994pv,Barbieri:1995tw} 
offers a sensitive probe of new physics and a nice possibility to study it 
experimentally providing an interesting interplay of particle and nuclear physics effects. 
The number of relevant operators in Eqs.~(\ref{eq:L_eff_6}) and (\ref{eq:L_eff}) is reduced if one only considers {\it coherent} 
$\mu+N \to e+N$ transitions\footnote{There are also important non-local contributions from the operators 
governing $\mu \to e \gamma$ transitions with the photon attached to a nucleus. Those contributions are well 
known~\cite{Czarnecki:1998iz} and will not be discussed here.}~\cite{Cirigliano:2009bz}. 

The initial state in the $\mu$-$e$ conversion process
\begin{eqnarray}
  \mu^- + (A,Z) \to e^- + (A,Z)^\prime
\end{eqnarray}
is the $1s$ state of the muonic atom with the binding energy $E_b$, and the final electron state is the eigenstate with the energy $m_\mu-E_b$ (neglecting the atomic recoil energy of a muonic atom, see~\cite{Bertl:2006up}). Following Ref.~\cite{Kitano:2002mt}  the $\mu$-$e$ conversion amplitude can be written as
\begin{eqnarray}\label{eq:amplitude}
     M_{NN^\prime}^{\mu e}   &=&    \frac{1}{\Lambda^2} \int d^3x  
  		 \left[ \left( c_1 \bar\psi_{\kappa,W}^{\mu(e)} P_L \psi_{1s}^{(\mu)}   \right.\right.  \nonumber\\
               &+&            \left.  c_3 \bar\psi_{\kappa,W}^{\mu(e)} P_R \psi_{1s}^{(\mu)} 
                          \right) \langle N^\prime |  \frac{\beta_L}{4\alpha_s}G_{\mu\nu}^a G^{a\mu\nu} | N \rangle  \nonumber\\
               &+&         \left( c_2 \bar\psi_{\kappa,W}^{\mu(e)} P_L \psi_{1s}^{(\mu)} 
                                + c_4 \bar\psi_{\kappa,W}^{\mu(e)} P_R \psi_{1s}^{(\mu)} 
                           \right) 		\nonumber\\
                &\times&   \left.     \langle N^\prime |  \frac{\beta_L}{4\alpha_s}G_{\mu\nu}^a \widetilde G^{a\mu\nu} | N \rangle \right], 
\end{eqnarray}
where $\langle N^\prime|$ and $|N\rangle$ are the final and initial states of the nucleus, respectively; 
the $1s$ initial muon wave function 
\begin{eqnarray}
  \psi_{1s}^{(\mu)}  =     \left( \begin{array}{c}
    g_\mu^- \chi_{-1}^{\pm1/2} \\ 
    \text{i} f_\mu^- \chi_1^{\pm1/2} \\ 
  \end{array} \right)
\end{eqnarray}
is normalized to 1 and corresponds to the quantum numbers $\kappa=-1$ 
 and $\mu=\pm1/2$ of the operators
 \begin{eqnarray}
 	K	=	\left(    \begin{array}{cc}
    	{\boldsymbol\sigma}\cdot{\boldsymbol{\it l}} +1 & 0 \\ 
    	0 & - ({\boldsymbol\sigma}\cdot{\boldsymbol{\it l}} +1) \\ 
  \end{array}	\right)
\end{eqnarray}
and $j_z$, respectively, where ${\boldsymbol{\it l}}$ is the orbital angular momentum;
and $\kappa=\pm1$ final electron wave functions
\begin{eqnarray}\label{eq:psi_e1}
  \psi_{-1,W}^{\pm1/2(e)}  =     \left( \begin{array}{c}
    g_e^- \chi_{-1}^{\pm1/2} \\ 
    \text{i} f_e^- \chi_1^{\pm1/2} \\ 
  \end{array} \right)
\end{eqnarray}
and
\begin{eqnarray}\label{eq:psi_e2}
  \psi_{1,W}^{\pm1/2(e)}  =     \left( \begin{array}{c}
    g_e^+ \chi_{1}^{\pm1/2} \\ 
    \text{i} f_e^+ \chi_{-1}^{\pm1/2} \\ 
  \end{array} \right)
\end{eqnarray}
are normalized as 
\begin{eqnarray}
 \int d^3x \  \psi_{\kappa,W}^{\mu(e)*}({\bf x})\psi_{\kappa^\prime,W^\prime}^{\mu^\prime(e)}({\bf x}) = 2\pi\delta_{\mu\mu^\prime}\delta_{\kappa\kappa^\prime}
 \delta(W-W^\prime),
\end{eqnarray}
where $W$ is the energy.
The electron mass was neglected in Eqs.~\eqref{eq:psi_e1} and \eqref{eq:psi_e2} so that $g_e^+=\text{i} f_e^-$ and $\text{i} f_e^+=g_e^-$. Using the normalization
\begin{eqnarray}
 		\int_{-1}^{1} d\cos\theta  \int_0^{2\pi} d\phi  \chi_\kappa^{\mu*} \chi_{\kappa^\prime}^{\mu^\prime} = \delta_{\mu\mu^\prime} \delta_{\kappa\kappa^\prime}
\end{eqnarray}
of the eigenfunctions $\chi_\kappa^\mu$ of $({\boldsymbol\sigma}\cdot{\boldsymbol{\it l}} +1)$ and $j_z$, we have
\begin{eqnarray}
 	\bar \psi_{-1,W}^{(e)}  P_\alpha \psi_{1s}^{(\mu)}  &=&   \frac{1}{2} \left( g_e^-g_\mu^- - f_e^- f_\mu^- \right)   \nonumber\\
											&=& \bar \psi_{1,W}^{(e)}  P_{R} \psi_{1s}^{(\mu)},  \\
	\bar \psi_{1,W}^{(e)}  P_{L} \psi_{1s}^{(\mu)}  &=&   -\frac{1}{2} \left( g_e^-g_\mu^- - f_e^- f_\mu^- \right)
\end{eqnarray}
with $\alpha = L,R$.

The pseudoscalar nucleon current couples to the nuclear spin leading to incoherent contribution~\cite{Kosmas:2001mv}. The matrix element in Eq.~\eqref{eq:amplitude} relevant to the coherent conversion process ($N=N^\prime$) can be expressed by the proton $\rho^{(p)}$ and neutron $\rho^{(n)}$ densities in a nucleus as
\begin{eqnarray}
	&& \langle N |  \frac{\beta_L}{4\alpha_s}G_{\mu\nu}^a G^{a\mu\nu} | N \rangle  	\nonumber\\
	 && =  -\frac{9}{2} \left[ Z G^{(g,p)} \rho^{(p)} + (A-Z) G^{(g,n)} \rho^{(n)} \right], 
\end{eqnarray}
where $A$ and $Z$ represent mass and atomic number of the nucleus, and the matrix element of gluon operator between the nucleon states is defined as
\begin{eqnarray}
   G^{(g,\mathcal{N})} = \langle \mathcal{N} |  \frac{\alpha_s}{4\pi} G_{\mu\nu}^a G^{a\mu\nu} | \mathcal{N} \rangle
\end{eqnarray}
with $\mathcal N=n,p$.
This scalar matrix element can be calculated by using the trace of the energy-momentum tensor $\Theta^\mu_{~\mu}$~\cite{Shifman:1978zn} and applying the flavor $SU(3)$ symmetry. 
Since $\Theta^\mu_{~\mu}$ is flavor symmetric, the proton and neutron scalar matrix elements are equal. 
For the strange-quark sigma term $\sigma_s \equiv m_s\langle p|\bar ss|p\rangle =50$~MeV the numerical result is $G^{(g,\mathcal{N})}  =  -189$~MeV~\cite{Cheng:2012qr}.

The nucleon densities are assumed spherically symmetric and normalized as
\begin{eqnarray}
   \int_0^\infty dr  4\pi r^2 \rho^{(\mathcal N)}(r) = 1.
\end{eqnarray}
As usual for spherical nuclei, the two-parameter Fermi (2pF) charge distribution is used
\begin{eqnarray}\label{eq:2pF_model}
   \rho(r) = \frac{\rho_0}{1+\exp[ (r-c) / z ]},
\end{eqnarray}
where $c$ is the half-density radius. 

The formula for the coherent conversion rate can be written as
\begin{eqnarray}
    \Gamma_\text{conv}(\mu N\to eN)  =   \frac{4}{\Lambda^4}   \left( |c_1|^2 + |c_3|^2 \right) a_N^2, 
\end{eqnarray}
where 
\begin{eqnarray}
  	a_N = G^{(g,p)} S^{(p)} + G^{(g,n)} S^{(n)}. 
\end{eqnarray}
The overlap integrals are defined as~\cite{Kitano:2002mt}
\begin{eqnarray}
   S^{(p)} = \frac{1}{2\sqrt{2}} \int_0^\infty dr r^2 Z \rho^{(p)} (g_e^-g_\mu^- - f_e^-f_\mu^-), \label{eq:Sp} \\
   S^{(n)} = \frac{1}{2\sqrt{2}} \int_0^\infty dr r^2 (A-Z) \rho^{(n)} (g_e^-g_\mu^- - f_e^-f_\mu^-).  \label{eq:Sn}
\end{eqnarray}
%

The parameters of model 2pF of nucleon densities in Eq.~\eqref{eq:2pF_model}~\cite{De Jager:1987qc}, and the overlap integrals in Eqs.~\eqref{eq:Sp} and \eqref{eq:Sn}~\cite{Kitano:2002mt} for the same distributions 
\begin{eqnarray}
   \rho^{(p)}(r)=\rho^{(n)}(r)\equiv \rho(r)
\end{eqnarray}
of neutrons and protons in the nuclei $^{48}_{22}$Ti and $^{197}_{79}$Au are shown in Table~\ref{Table:several_nuclei}. 
\begin{table}[htp]
\caption{Nucleon densities model parameters, and the overlap integrals in the unit of $m_\mu^{5/2}$ for several nuclei}
\begin{center}
\begin{tabular}{c|c|c|c|c|c}
\hline
\hline
	Nucleus 			& Model	& $c$,~fm 	& $z$,~fm		& $S^{(p)}$	& $S^{(n)}$  \\ 
	\hline
	$^{48}_{22}$Ti		&	FB	&	$-$		&	$-$		& 0.0368		& 0.0435	 \\ 
	\hline
	$^{197}_{79}$Au	&	2pF	& 6.38		& 0.535		& 0.0614		& 0.0918	 \\ 
\hline
\hline
\end{tabular}
\end{center}
\label{Table:several_nuclei}
\end{table}
The parameters of the Fourier-Bessel expansion (FB)
\begin{eqnarray}
  	\rho(r) = \left\{     \begin{array}{ccc}
    \sum\limits_v a_v\sin(v\pi rR^{-1})/(v\pi rR^{-1})	\ 	&\text{for}&	\ 	r\leq R  \\ 
    0								\ 	&\text{for}&	\ 	r > R \\ 
  \end{array} \right.
\end{eqnarray}
are given in Ref.~\cite{De Jager:1987qc,Fricke:1995zz}.

The branching ratio for $\mu$-$e$ conversion on a nucleus $N$ is defined as
\begin{eqnarray}
   B_{\mu e}^N \equiv \Gamma_\text{conv}(\mu^- N \to e^-N_{\rm g.s.}) / \Gamma_{\rm capt}(\mu^-{N}), 
\end{eqnarray}
where g.s. stands for ground state.
The SM muon capture rates~\cite{Suzuki:1987jf,Kitano:2002mt}, the upper bounds on $B_{\mu e}^{N}$, and the binding energies of $^{48}_{22}$Ti and $^{197}_{79}$Au are given in Table~\ref{Table:several_nuclei_bounds} with the respective references.
\begin{table}[htp]
\caption{Muon capture rates, bounds on $B_{\mu e}^N$, and binding energies for several nuclei $N$}
\begin{center}
\begin{tabular}{c|c|c|c}
\hline
\hline
	Nucleus 		 & $\Gamma_\text{capt}(\mu^-N)$,~$\text{s}^{-1}$ & $B_{\mu e}^N$~$(90\%~\text{C.L.})$							&	$E_b$, MeV  \\ 
	\hline
	$^{48}_{22}$Ti	 & $2.59\times10^6$							& $4.3\times10^{-12}$ Ref.~\cite{Dohmen:1993mp}		&	1.25 Ref.~\cite{Kosmas:1997eca} \\ 
	\hline
	$^{197}_{79}$Au & $13.07\times10^6$						& $7\times10^{-13}$ Ref.~\cite{Bertl:2006up} &	10.08 Ref.~\cite{Bertl:2006up} \\ 
\hline	
\hline
\end{tabular}
\end{center}
\label{Table:several_nuclei_bounds}
\end{table}
The upper bounds on the parameters of the Lagrangian in Eq.~\eqref{eq:L_eff} for one nonzero coefficient $c$ at a time are given in 
Table~\ref{Table:3}. It shows that the bound on $\mu$-$e$ conversion rate on gold gives the best 
limit (see also Ref.~\cite{Harnik:2012pb}).
%
\begin{table}[tp]
\caption{Upper bounds on the parameters of the Lagrangian in Eq.~\eqref{eq:L_eff} 
from muon conversion experiments.}
\begin{center}
\begin{tabular}{c|c|c}
\hline
\hline
 %
Coefficient &  \multicolumn{2}{c}{ $~$Bound on $|c_i^{e\mu}|/\Lambda^2$, GeV$^{-3}$ $~$}  \\ 
  \hhline{~|-|-}
  	&  conversion on  $^{48}_{22}$Ti	&	 conversion on  $^{197}_{79}$Au	 \\
\hline
\hline
  $c_1$ &  $2.5\times 10^{-11}$  &  $1.2\times 10^{-11}$   \\ 
  $c_2$ &  $-$  &  $-$   \\ 
  $c_3$ &  $2.5\times 10^{-11}$  &  $1.2\times 10^{-11}$   \\ 
  $c_4$ &  $-$  &  $-$   \\ 
\hline
\hline
\end{tabular}
\end{center}
\label{Table:3}
\end{table}

\section{Constraints from $\tau$ decays.}\label{Ch:TauDecays}

The analysis presented above only deals with experimentally-interesting coherent $\mu-e$ conversion. As a result, no
parity-violating operator give any contribution to the experimental transition rates. Moreover, we had to resort to 
models to describe nuclear effects affecting conversion rates. It might be advantageous to 
use other experimental observables to study LFV new physics couplings to heavy quarks via gluonic
operators. LFV tau decays offer such opportunity. Besides, analyses of tau decays have different theoretical uncertainties 
than muon conversion calculations; in fact, one can use chiral symmetry and low energy theorems to provide 
model-independent evaluations of operator matrix elements. 
While the tau decays have been studied in a variety of models~\cite{Li:2005rr}, to the best of our knowledge 
gluonic operator contribution (and thus constraints on heavy quark couplings from those decays) has not been previously 
considered. In what follows we shall use tau decays to constrain matrix elements of gluonic operators.

\subsection{Parity-conserving gluonic operators}

Complimentary to muon conversion experiments considered in section~\ref{Ch:Conversion}, parity-conserving operators 
can also be probed in lepton-flavor-violating tau decays $\tau \to \ell M^+M^-$, where $\ell = \mu, e$ and 
$M=\pi,\eta^{(\prime)}, K$~\cite{Black:2002wh,Chen:2006hp}.
In what follows, let us concentrate on the case of three-body decays $\tau \to \mu \pi^+\pi^-$ and 
$\tau \to \mu K^+ K^-$, which are the most interesting experimentally since all the final particles are charged. 
Transitions to other states (like $\tau \to \ell \eta\eta$) can be obtained by employing flavor $SU(3)$ symmetry relations.

In order to constrain the Wilson coefficients of the operators in Eq.~(\ref{Eq:ScalOps}), Eq.~(\ref{Eq:VecOps}), and 
Eq.~(\ref{eq:Oprs}) one needs to constrain hadronic matrix elements. For the scalar operators we shall follow~\cite{Black:2002wh} 
to state
\begin{eqnarray}\label{Eq:ScalMatrEl}
\langle \pi^+\pi^- | \bar q q | 0 \rangle= \langle K^+K^- | \bar q q | 0 \rangle=  \delta^M_q B_0,
\end{eqnarray}
where for charged final states $\delta^M_q=1$ if the flavor of the quark field $q$ in the operator matches the flavor content of the 
meson and zero otherwise. For example,  $\langle K^+K^- | \bar s s | 0 \rangle= \langle K^+K^- | \bar u u | 0 \rangle= B_0$, while 
$\langle K^+K^- | \bar d d | 0 \rangle=  0$. Matrix elements for other light final states can be related to Eq.~(\ref{Eq:ScalMatrEl})
via flavor $SU(3)$ relations~\cite{Black:2002wh}, e.g.,
\begin{eqnarray}
3 \langle \eta_8 \eta_8 | \bar u u | 0 \rangle= \frac{3}{4} \langle  \eta_8 \eta_8 | \bar q q | 0 \rangle=  B_0.
\end{eqnarray}
Note that $B_0 = 1.96$~GeV can be estimated from the chiral Lagrangian relations, $m_\pi^2 = \left(m_u+m_d\right) B_0$
assuming that $m_u=m_d=5$~MeV.

For the vector operators Eq.~(\ref{Eq:VecOps}) one can use the definition of the pion (kaon) form factor and 
crossing symmetry,
\begin{eqnarray}\label{Eq:VecMatrEl}
\langle M^+M^- | \bar q \gamma_\mu q | 0 \rangle = \delta^M_q G_M^{(q)} (Q^2)  \left(p_+-p_-\right)_\mu,
\end{eqnarray}
where $Q=p_\tau-p_\ell$ is the momentum transfer to the hadronic system and $p_\pm$ are the 4-momenta of $M^\pm$.
Note that $G_M^{(q)} (0)=1$~\cite{Daub:2012mu}. Just like before, flavor content
of the operators should match that of the final state mesons.

The matrix elements of gluonic operators Eq.~(\ref{eq:Oprs}) are easiest estimated in the chiral limit, where $m_u=m_d=m_s=m_M=0$. 
In that limit, a low-energy theorem states that~\cite{Voloshin:1985tc}
\begin{eqnarray}
\langle M^+M^- |  \frac{\alpha_s} {4\pi}G^{a\mu\nu} G_{\mu\nu}^a | 0 \rangle = - \frac{2}{9} q^2,
\end{eqnarray}
We do not expect the results to change 
much away from the chiral limit, so shall use this estimate in what follows. 
It is these operators that we are most interested in the paper.
Finally, parity invariance of strong interactions implies that
\begin{eqnarray}
\langle M^+M^- | \bar q \gamma_5 q | 0 \rangle 
&=& \langle M^+M^- | \frac{\alpha_s}{4\pi}G^{a\mu\nu} \widetilde G_{\mu\nu}^a | 0 \rangle 
\nonumber \\
&=& \langle M^+M^- | \bar q \gamma_\mu \gamma_5 q | 0 \rangle = 0.
\end{eqnarray}
With the definitions above one can calculate the differential decay rate for the decay $\tau \to \ell M^+ M^-$. For the scalar 
and gluonic operators one has
\begin{eqnarray}\label{Eq:DiffDecay}
\frac{d \Gamma (\tau \to \ell M^+ M^-)}{d q^2} &=&
\frac{m_\tau}{32 (2 \pi)^3 \Lambda^4} 
\left[
\left|A_{MM}\right|^2 + \left|B_{MM}\right|^2
\right]
\nonumber \\
&\times&
\sqrt{1-\frac{4 m_M^2}{q^2}} 
\left(1-\frac{q^2}{m_\tau^2}\right)^2,
\end{eqnarray}
where we set $m_\ell=0$ and defined 
\begin{eqnarray}
A_{MM} &=& - \frac{2 c_1^{\ell\tau}}{9} q^2 + \frac{1}{2}
\sum_{q=u,d,s} \left(C_1^{q\ell\tau}+C_2^{q\ell\tau}\right) \delta^M_q B_0,
\nonumber \\
B_{MM} &=& - \frac{2 c_3^{\ell\tau}}{9} q^2 + \frac{1}{2}
\sum_{q=u,d,s} \left(C_3^{q\ell\tau}+C_4^{q\ell\tau}\right)\delta^M_q B_0.
~~~~
\end{eqnarray}
Integrating Eq.~(\ref{Eq:DiffDecay}) we obtain constraints on $c_1^{\ell\tau}$ and $c_3^{\ell\tau}$ listed in Table~\ref{Table:3a}.
Finally, for completion, we present the contribution to the differential decay rate due to vector operators,
\begin{eqnarray}\label{Eq:DiffDecayVec}
&&\frac{d \Gamma_V (\tau \to \ell M^+ M^-)}{d q^2} =
\frac{m_\tau^3}{768 \pi^3 \Lambda^4} 
\left[
\left|C_{MM}\right|^2 + \left|D_{MM}\right|^2
\right]
\nonumber \\
&&\qquad \times
\left(1-\frac{4 m_M^2}{q^2}\right)^{3/2}
\left(1-\frac{q^2}{m_\tau^2}\right)^2
\left(1-2\frac{q^2}{m_\tau^2}\right),
\end{eqnarray}
where we set $m_\ell=0$ and defined 
\begin{eqnarray}
C_{MM} &=&  \frac{1}{2}
\sum_{q=u,d,s} \left(C_5^{q\ell\tau}+C_6^{q\ell\tau}\right) \delta^M_q G_M
\nonumber \\
D_{MM} &=&  \frac{1}{2}
\sum_{q=u,d,s} \left(C_7^{q\ell\tau}+C_8^{q\ell\tau}\right)\delta^M_q G_M.
~~~~
\end{eqnarray}
This result could be used to constrain Wilson coefficients of vector operators.

%
\begin{table*}[tp]
\caption{Upper bounds on the parameters of the Lagrangian in Eq.~\eqref{eq:L_eff} 
from tau decay experiments.}
\begin{center}
\begin{tabular}{c|c|c|c|c|c|c|c|c}
\hline
\hline
&  \multicolumn{8}{c}{ $~$Bound on $|c_i^{\ell\tau}|/\Lambda^2$, GeV$^{-3}$ $~$}  \\ 
  \hhline{~|-|-|-|-|-|-|-|-}
 Coef $~$     &  ${\cal B} (\tau \to \mu \ \pi^+\pi^-)$	 & ${\cal B} (\tau \to e \ \pi^+\pi^-)$	&  ${\cal B} (\tau \to \mu K^+K^-)$ & ${\cal B} (\tau \to e K^+K^-)$ &  
          ${\cal B} (\tau \to \mu \eta^{\prime})$ & ${\cal B} (\tau \to e \eta^{\prime})$ &  
          ${\cal B} (\tau \to \mu \eta)$ & ${\cal B} (\tau \to e \eta)$ \\
      &  $< 2.1\times10^{-8}$	 & $< 2.3\times10^{-8}$ &  $< 4.4\times10^{-8}$ & $< 3.3 \times10^{-8}$ &  
          $<1.3\times10^{-7}$ & $< 1.6\times10^{-7}$ &  
          $< 1.3\times10^{-7}$ & $< 1.6\times10^{-7}$ \\   
\hline
\hline
  $c_1$ &  $6.8\times10^{-8}$  &  $6.5\times10^{-8}$   &  $9.4\times10^{-8}$  &  $8.2\times10^{-8}$ &  $-$  &  $-$ &  $-$  &  $-$\\ 
  $c_2$ &  $-$  &  $-$   &  $-$  &  $-$ &  $2.3\times10^{-7}$  &  $2.5\times10^{-7}$ &  $1.6\times10^{-7}$  &  $1.5\times10^{-7}$\\ 
  $c_3$ &  $6.8\times10^{-8}$  &  $6.5\times10^{-8}$   &  $9.4\times10^{-8}$  &  $8.2\times10^{-8}$ &  $-$  &  $-$ &  $-$  &  $-$\\ 
  $c_4$ &  $-$  &  $-$   &  $-$  &  $-$ & $2.3\times10^{-7}$  &  $2.5\times10^{-7}$ &  $1.6\times10^{-7}$  &  $1.5\times10^{-7}$\\ 
\hline
\hline
\end{tabular}
\end{center}
\label{Table:3a}
\end{table*}
%

\subsection{Parity-violating gluonic operators}

Constraints on the parity-violating contributions can be obtained from the lepton-flavor-violating meson and tau decays, 
$\tau \to \ell M$ and $M \to \mu e$, where $\ell = \mu, e$, and $M=\pi,\eta, \eta^\prime$. The analysis of decays involving 
an $\eta^\prime$ is especially interesting, as $\eta^\prime$ meson contains considerable amount of ``glue'', which makes it possible 
to constrain gluonic LFV operators resulting from integrating out heavy quarks. 

To calculate the decay rates one needs to parameterize the hadronic matrix elements,
\begin{eqnarray}\label{DecConstants}
&& \langle M (p) |  \bar q \gamma^\mu \gamma_5 q | 0 \rangle = - i b_q f_M^q p^\mu,
\nonumber \\
&& \langle M (p) |  \bar q  \gamma_5 q | 0 \rangle = -i b_q h_M^q,
\\
&& \langle M (p) |  \frac{\alpha_s}{4\pi} G^{a\mu\nu} \widetilde G^a_{\mu\nu} | 0 \rangle = a_M,
\nonumber
\end{eqnarray}
where $q=u, d, s$, and $b_{u,d}=1/\sqrt{2}$, $b_s = 1$. The form factors defined above in the
Feldmann-Kroll-Stech (FKS) mixing scheme~\cite{Feldmann:1998vh} are constrained for $\eta$ and $\eta^\prime$ 
mesons to be~\cite{Beneke:2002jn}
\begin{eqnarray}
a_\eta &=& -\frac{m_{\eta^\prime}^2-m_\eta^2}{2} \sin2\phi 
\left(-f_q b_q \sin\phi+f_s\cos\phi\right), 
\nonumber \\
a_{\eta^\prime} &=& -\frac{m_{\eta^\prime}^2-m_\eta^2}{2} \sin2\phi 
\left(f_q b_q \sin\phi+f_s\cos\phi\right),
\end{eqnarray}
where $\phi=39.3^o\pm1.0^o$ is a mixing angle of $\eta$ and $\eta^\prime$ in the flavor basis~\cite{Beneke:2002jn}. Numerically,
anomaly matrix elements are $a_\eta=-0.022\pm 0.002$~GeV$^3$, $a_{\eta'}=-0.057\pm 0.002$~GeV$^3$. The decays constants in 
Eq.~(\ref{DecConstants}) used in numerical work are $f^q_\eta = 108\pm 3$~MeV, $f^q_{\eta^\prime} = 89\pm 3$~MeV,
$f^s_\eta = -111\pm 6$~MeV, and $f^s_{\eta^\prime} = 136\pm 6$~MeV~\cite{Beneke:2002jn,Petrov:1997yf}.

Neglecting terms of the order ${\cal O}(m_\ell/m_\tau)$, which would change our answer to at most $5\%$ for $\ell=\mu$, we can write 
for the decay rate,
\begin{equation}
\Gamma \left( \tau \to \mu M\right ) = \frac{m_\tau}{8\pi \Lambda^4}
\left[ 
\left|A_M\right|^2 +  \left|B_M\right|^2
\right] 
\left(1-\frac{m_M^2}{m_\tau^2}\right)^2,
\end{equation}
where $A_M$ and $B_M$ are defined as
\begin{eqnarray}
A_M = -\frac{2i}{9} c_2^{\ell\tau} a_M &+& \sum_{q=u,d,s} \left(C_2^{q\ell\tau}-C_1^{q\ell\tau}\right) \frac{b_q h^q_M}{4 m_q} 
\nonumber \\
&+& \frac{1}{2} m_\mu \sum_{q=u,d,s} \left(C_5^{q\ell\tau}-C_6^{q\ell\tau}\right) b_q f^q_M ~~~~~
\\
&-& \frac{1}{2} m_\tau \sum_{q=u,d,s} \left(C_7^{q\ell\tau}-C_8^{q\ell\tau}\right) b_q f^q_M,
\nonumber \\
B_M =  -\frac{2i}{9} c_4^{\ell\tau} a_M &+& \sum_{q=u,d,s} \left(C_4^{q\ell\tau}-C_3^{q\ell\tau}\right) \frac{b_q h^q_M}{4 m_q} 
\nonumber \\
&-& \frac{1}{2} m_\tau \sum_{q=u,d,s} \left(C_5^{q\ell\tau}-C_6^{q\ell\tau}\right) b_q f^q_M ~~~~
\\
&+& \frac{1}{2} m_\mu \sum_{q=u,d,s} \left(C_7^{q\ell\tau}-C_8^{q\ell\tau}\right) b_q f^q_M.
\nonumber 
\end{eqnarray}
The current experimental bounds on flavor-violating tau decays allow to put stringent constraints on 
Wilson coefficients $c_i^{\ell\tau}$~\cite{Beringer:1900zz}. We display them in Table~\ref{Table:3a}.
These results, along with the ones displayed in Table~\ref{Table:3}, can be translated into bounds on 
flavor-changing interactions of leptons with heavy quarks in particular models. As an example on
how this can be done, we consider a generic leptoquark model.

\section{Model example: leptoquarks}\label{Ch:Leptoquarks}

The Wilson coefficients of effective gluonic operators that were constrained in the previous sections can be used to put
bounds on parameters of operators describing lepton interactions with heavy quarks in particular models of NP. Let us
provide an example of how this can be done using generic leptoquark model. 

The general renormalizable, $B$ and $L$ conserving, and $SU(3)\times SU(2)\times U(1)$ invariant LQ-lepton-quark interactions are given in Refs.~\cite{Buchmuller:1986zs,Davidson:1993qk,Gonderinger:2010yn}.
The relevant for our considertion scalar LQs ($S$) and vector LQs ($V$) interactions are
\begin{eqnarray}\label{eq:LQ-l-q}
	   \mathcal{L}_{S}  &=&  \left(  \lambda_{LS_0} \bar q_L^c \text{i}\tau_2 \ell_{2L}  +  \lambda_{RS_0} \bar u_R^c \ell_{1R}  \right)  S_0^\dag 	\nonumber\\
	  &+& \left(  \lambda_{LS_{1/2}} \bar u_R \ell_{2L}  +  \lambda_{RS_{1/2}} \bar q_L \text{i}\tau_2 \ell_{1R}  \right)  S_{1/2}^\dag + \text{H.c.}, ~~ \\
	   \mathcal{L}_{V}  &=&  \left(  \lambda_{LV_0} \bar q_L \gamma_\mu \ell_{2L}  +  \lambda_{RV_0} \bar d_R \gamma_\mu \ell_{1R}  \right)  V_0^{\mu\dag} 	\nonumber\\
	  &+& \left(  \lambda_{LV_{1/2}} \bar d_R^c \gamma_\mu \ell_{2L}  +  \lambda_{RV_{1/2}} \bar q_L^c \gamma_\mu \ell_{1R}  \right)  V_{1/2}^{\mu\dag} + \text{H.c.},     \label{eq:LQ-l-q:2}\nonumber \\
\end{eqnarray}
where $q$, $u$ and $d$ are doublet, singlet up and singlet down quarks, respectively;  we omit flavor indexes, the subindexes 0 and 1/2 indicate $SU(2)$ singlet and doublet LQ, respectively;  and couplings $\lambda$ are assumed to be real.

Consider $\mu$-$e$ conversion on $^{197}_{79}$Au induced by leptoquark exchange. 
For the values of the loop integral in Eq.~\eqref{eq:loop_integral1} we simply have $I_1(m_t)=I_1(m_b)=I_1(m_c)=0.333$ since the muon mass and the 
binding energy of the muonic gold are much lower than $c$, $b$ and $t$ quark masses. 
The expressions for relevant Wilson coefficients in Eq.~\eqref{eq:L_eff_6} are given in Table~\ref{Table:4}, where the quark flavor indexes are 
$u=u,c,t$ and $d=d,s,b$.
%
%
\begin{table}[htp]
\caption{The Wilson coefficients for the model with LQs}
\begin{center}
\begin{tabular}{c|c||c|c}
		\hline\hline
			$C_i^u/\Lambda^2$			&	Expression	&	$C_i^d/\Lambda^2$	&	Expression	\\
		\hline
			$\frac{C_1^u}{\Lambda^2}$	&	$\frac{\lambda_{RS_{1/2}}^{\ell_1u}\lambda_{LS_{1/2}}^{\ell_2 u}}{2M_{S_{1/2}}^2}$	&	$\frac{C_1^d}{\Lambda^2}$			&	$\frac{\lambda_{LV_{1/2}}^{\ell_2 b}\lambda_{RV_{1/2}}^{\ell_1 b}}{M_{V_{1/2}}^2}$	\\
		\hline
			$\frac{C_2^u}{\Lambda^2}$	&	$\frac{\lambda_{RS_0}^{\ell_1 u}\lambda_{LS_0}^{\ell_2 u}}{2M_{S_0}^2}$	&	$ \frac{C_2^d}{\Lambda^2}$	&	
			$\frac{\lambda_{LV_0}^{\ell_2 b}\lambda_{RV_0}^{\ell_1 b}}{M_{V_0}^2}$	\\
		\hline
			$\frac{C_3^u}{\Lambda^2}$	&	$\frac{\lambda_{RS_0}^{\ell_2 u}\lambda_{LS_0}^{\ell_1 u}}{2M_{S_0}^2}$	&	$		\frac{C_3^d}{\Lambda^2}$			&	$\frac{\lambda_{LV_0}^{\ell_1 b}\lambda_{RV_0}^{\ell_2 b}}{M_{V_0}^2}$	\\
		\hline
			$\frac{C_4^u}{\Lambda^2}$	&	$\frac{\lambda_{RS_{1/2}}^{\ell_2 u}\lambda_{LS_{1/2}}^{\ell_1 u}}{2M_{S_{1/2}}^2}$	&	$\frac{C_4^d}{\Lambda^2}$			&	$\frac{\lambda_{LV_{1/2}}^{\ell_1 b}\lambda_{RV_{1/2}}^{\ell_2 b}}{M_{V_{1/2}}^2}$		\\
		\hline\hline
\end{tabular}
\end{center}
\label{Table:4}
\end{table}
%
%

We assume that only the couplings $\lambda$ for a single quark flavor are nonzero at a time. 
From Eqs.~\eqref{eq:c1} and \eqref{eq:c3} it follows that the best bounds for the scalar LQs (left half of the Table~\ref{Table:4}) are 
relaxed by the factor $2m_t/m_b\simeq 75$ with respect to the ones for the vector LQs (right half of the Table~\ref{Table:4}). 
Using the bound on $|c_1|$, we have for $e=\ell_1$ and $\mu=\ell_2$
\begin{eqnarray}
    	\frac{|\lambda^{et}_{RS_0}\lambda^{\mu t}_{LS_0}|}{M_{S_0}^2} = \frac{|\lambda^{et}_{RS_{1/2}}\lambda^{\mu t}_{LS_{1/2}}|}{M_{S_{1/2}}^2}  <  1.2\times10^{-8}~\text{GeV}^{-2},   \\  
   	\frac{|\lambda^{\mu b}_{LV_0}\lambda^{eb}_{RV_0}|}{M_{V_0}^2} = \frac{|\lambda^{\mu b}_{LV_{1/2}}\lambda^{eb}_{RV_{1/2}}|}{M_{V_{1/2}}^2}  <  1.6\times10^{-10}~\text{GeV}^{-2},
\end{eqnarray}
and, using the bound on $|c_3|$, we have
\begin{eqnarray}
    	\frac{|\lambda^{\mu t}_{RS_0}\lambda^{et}_{LS_0}|}{M_{S_0}^2} = \frac{|\lambda^{\mu t}_{RS_{1/2}}\lambda^{et}_{LS_{1/2}}|}{M_{S_{1/2}}^2}  <  1.2\times10^{-8}~\text{GeV}^{-2},   \\  
   	\frac{|\lambda^{eb}_{LV_0}\lambda^{\mu b}_{RV_0}|}{M_{V_0}^2} = \frac{|\lambda^{eb}_{LV_{1/2}}\lambda^{\mu b}_{RV_{1/2}}|}{M_{V_{1/2}}^2}  <  1.6\times10^{-10}~\text{GeV}^{-2}.
\end{eqnarray}
Finally, for the common scales $M_S$ and $M_V$ of scalar and vector LQ masses, respectively, we get
\begin{eqnarray}
    	|\lambda^{\alpha t}_{RS_0}\lambda^{\beta t}_{LS_0}| 
    	= |\lambda^{\alpha t}_{RS_{1/2}}\lambda^{\beta t}_{LS_{1/2}}|  
    	<  1.2\times10^{-2}~\left( \frac{M_{S}}{1~\text{TeV}} \right)^2,   \\  
   	|\lambda^{\alpha b}_{LV_0}\lambda^{\beta b}_{RV_0}| 
   	= |\lambda^{\alpha b}_{LV_{1/2}}\lambda^{\beta b}_{RV_{1/2}}|  
   	<  1.6\times10^{-4}~\left( \frac{M_{V}}{1~\text{TeV}} \right)^2,
\end{eqnarray}
where $\alpha\neq\beta=e,\mu$. 

In leptoquark models the couplings of heavy quarks can also be constrained from the photon dipole-like operators that also contribute to $\mu \to e \gamma$. Those have been recently
constrained in Ref.~\cite{Gabrielli:2000te}. Assuming the dominance of the dipole operator over all other contributions, one obtains comparable bounds 
on heavy quark couplings to leptoquarks which are of order $10^{-3}(M_{LQ}/1\,{\rm TeV})^2$ for the products of couplings with same chiralities 
$|\lambda_{LQ}^{\mu i}\lambda_{LQ}^{ei}|$. Here we only considered quarks of the last two generations $i=2,3$ (either $c,t$ or $s,b$), 
$LQ = LS_0$, $RS_0$, $LS_{1/2}$, $RS_{1/2}$, $LV_0$, $RV_0$, $LV_{1/2}$ and $RV_{1/2}$, with $M_{LQ}$ being the mass of the correspondent scalar or 
vector LQ~\cite{Gabrielli:2000te}.

Similar constraints are also available from tau decays. For $\mu=\ell_1$ and $\tau=\ell_2$
\begin{eqnarray}
    	\frac{|\lambda^{\mu t}_{RS_0}\lambda^{\tau t}_{LS_0}|}{M_{S_0}^2} = \frac{|\lambda^{\mu t}_{RS_{1/2}}\lambda^{\tau t}_{LS_{1/2}}|}{M_{S_{1/2}}^2}  <  2.3\times10^{-4}~\text{GeV}^{-2},   \\  
   	\frac{|\lambda^{\tau b}_{LV_0}\lambda^{\mu b}_{RV_0}|}{M_{V_0}^2} = \frac{|\lambda^{\tau b}_{LV_{1/2}}\lambda^{\mu b}_{RV_{1/2}}|}{M_{V_{1/2}}^2}  <  4.4\times10^{-6}~\text{GeV}^{-2},
\end{eqnarray}
and, 
\begin{eqnarray}
    	\frac{|\lambda^{\tau t}_{RS_0}\lambda^{\mu t}_{LS_0}|}{M_{S_0}^2} = \frac{|\lambda^{\tau t}_{RS_{1/2}}\lambda^{\mu t}_{LS_{1/2}}|}{M_{S_{1/2}}^2}  <  2.3\times10^{-4}~\text{GeV}^{-2},   \\  
   	\frac{|\lambda^{\mu b}_{LV_0}\lambda^{\tau b}_{RV_0}|}{M_{V_0}^2} = \frac{|\lambda^{\mu b}_{LV_{1/2}}\lambda^{\tau b}_{RV_{1/2}}|}{M_{V_{1/2}}^2}  <  4.4\times10^{-6}~\text{GeV}^{-2}.
\end{eqnarray}
While for $e=\ell_1$ and $\tau=\ell_2$,
\begin{eqnarray}
    	\frac{|\lambda^{et}_{RS_0}\lambda^{\tau t}_{LS_0}|}{M_{S_0}^2} = \frac{|\lambda^{et}_{RS_{1/2}}\lambda^{\tau t}_{LS_{1/2}}|}{M_{S_{1/2}}^2}  <  2.2\times10^{-4}~\text{GeV}^{-2},   \\  
   	\frac{|\lambda^{\tau b}_{LV_0}\lambda^{eb}_{RV_0}|}{M_{V_0}^2} = \frac{|\lambda^{\tau b}_{LV_{1/2}}\lambda^{eb}_{RV_{1/2}}|}{M_{V_{1/2}}^2}  <  4.2\times10^{-6}~\text{GeV}^{-2},
\end{eqnarray}
and, 
\begin{eqnarray}
    	\frac{|\lambda^{\tau t}_{RS_0}\lambda^{et}_{LS_0}|}{M_{S_0}^2} = \frac{|\lambda^{\tau t}_{RS_{1/2}}\lambda^{et}_{LS_{1/2}}|}{M_{S_{1/2}}^2}  <  2.2\times10^{-4}~\text{GeV}^{-2},   \\  
   	\frac{|\lambda^{eb}_{LV_0}\lambda^{\tau b}_{RV_0}|}{M_{V_0}^2} = \frac{|\lambda^{eb}_{LV_{1/2}}\lambda^{\tau b}_{RV_{1/2}}|}{M_{V_{1/2}}^2}  <  4.2\times10^{-6}~\text{GeV}^{-2}.
\end{eqnarray}
Clearly, constraints on the coefficients of operators containing tau-lepton fields are much weaker than the ones containing muon 
fields. We expect those constraints to improve with new data coming 
from Belle II collaboration.

\section{Conclusions}\label{Ch:Conclusions}

We considered contributions of heavy quark-induced operators to leptonic FCNC transitions. We constrained Wilson coefficients of 
effective gluonic operators in $\mu-e$, $\tau-\mu$, and $\tau-e$ transitions. These bounds can be used to study interactions of 
leptonic FCNCs with heavy quarks that are kinematically inaccessible in the described experiments. We provided an explicit example 
of constraints on the parameters of a generic leptoquark model.

\acknowledgments

We thank Will Shepherd for useful conversations, and German Valencia for pointing out a computational 
error in Table~\ref{Table:3}. 
A.A.P. would like to thank Kavli Institute for Theoretical Physics at
the University of California Santa Barbara for hospitality where part of this work was performed.
This work was supported in part by the U.S.\ Department of Energy under contract DE-FG02-12ER41825 and
by the National Science Foundation under Grant No. NSF PHY11-25915.

\appendix
\section{Integrals}\label{Ap:Integrals}

In this appendix we present the results of computation of the integrals needed to obtain the matching coefficients of
gluonic operators of dimension 7. The matching coefficients can be computed to be
\begin{eqnarray}
 	I_1(m_q)	=	\lambda_q \int_0^1 dx	\int_0^{1-x} dy	\frac{1-4xy}{\lambda_q-xy},  \label{eq:loop_integral1}	\\
	I_2(m_q)	=	\lambda_q \int_0^1 dx	\int_0^{1-x} dy	\frac{1}{\lambda_q-xy},
\end{eqnarray}
where $\lambda_q = m_q^2/E^2$ with the process energy scale $E$. 
For $\lambda_q >1/4$ they take the form
\begin{eqnarray}
	I_1 &=&   2\lambda_q -  \lambda_q(4\lambda_q-1)  	
	 \left[  \text{Li}_2\left(\frac{1+\text{i}\sqrt{-1+4\lambda_q}}{2\lambda_q}\right) \right.	\\
	&+& \left. \text{Li}_2\left(-\frac{2\text{i}}{\sqrt{-1+4\lambda_q}-\text{i}}\right)  \right],	\\
	I_2 &=&   \lambda_q  \left[  \text{Li}_2\left(\frac{1+\text{i}\sqrt{-1+4\lambda_q}}{2\lambda_q}\right) \right.	\\
	&+& \left. \text{Li}_2\left(-\frac{2\text{i}}{\sqrt{-1+4\lambda_q}-\text{i}}\right)  \right],
\end{eqnarray}
for $\lambda_q \gg 1$ they are
\begin{eqnarray}\label{eq:I1result_full}
	I_1 =   \frac{1}{3}\left[ 1  +  \frac{7}{120} \lambda_q^{-1}   +  \mathcal{O}(\lambda_q^{-2}) \right],	\\
	I_2 =   \frac{1}{2}\left[ 1  +  \frac{1}{12} \lambda_q^{-1}   +  \mathcal{O}(\lambda_q^{-2}) \right].
\end{eqnarray}
In this paper we use the leading order result in the $\lambda_q$-expansion.



\begin{thebibliography}{99}
\bibitem{ScalarOp}
For example, $Q_1$ operator would result 
from an operator of the form  $(\bar \ell_{1R} H L_{2L}) \ (\bar q_R H Q_L)$ (where $L_{2L}$ and $Q_L$ are the 
correspondingly electroweak doublets of leptons and quarks). This is a dimension 8 operator that is 
suppressed by two powers of the NP scale and two powers of the scale $v$ associated with electroweak 
symmetry breaking.
\end{thebibliography}
\end{document}